\documentclass[twocolumn,pof,aps,showpacs,preprintnumbers,amsmath,amssymb]{revtex4}

\usepackage{graphicx}
\usepackage{dcolumn}
\usepackage{bm}

\input epsf

\newcommand{\uvc}[1]{\bm{\mathrm{\hat #1}}} 
\newcommand{\mig}{{\mathrm{mig}}}
\newcommand{\cd}{0}
\newcommand{\surf}{s}
\newcommand{\bv}{{\bf v}}
\newcommand{\bu}{{\bf u}}

\newcommand{\by}{{\bf y}}

\newcommand{\br}{{\bf r}}

\newcommand{\Ca}{\mbox{\it Ca}}
\newcommand{\Ma}{\mbox{\it Ma}}

\newcommand{\bU}{{\bf U}}

\newcommand{\im}{{\mathrm i}}
\newcommand{\refeq}[1]{(\ref{#1})}
\newcommand{\eq}{{\mathrm{eq}}}
\newcommand{\visrat}{{\lambda}}
\newcommand{\dif}{{\mathrm  d}}
\newcommand{\ins}{{\mathrm  {in}}}
\newcommand{\out}{{\mathrm  {out}}}
\newcommand{\bt}{{\bf t}}

\begin{document}

\title{Surfactant-induced migration of a spherical drop in Stokes flow}
\author{James A. Hanna and Petia M. Vlahovska }
\affiliation{Thayer School of Engineering, Dartmouth College, 8000 Cummings Hall, Hanover, NH 03755, USA}
\date{\today}

\begin{abstract}
In Stokes flows, symmetry considerations dictate that a neutrally-buoyant
spherical particle will not migrate laterally with respect to the local flow
direction. We show that a loss of symmetry due to flow-induced surfactant
redistribution leads to cross-stream drift of a spherical drop in Poiseuille
flow.  We derive analytical expressions for the migration velocity in the limit
of small non-uniformities in the surfactant distribution, corresponding to
weak-flow conditions or a high-viscosity drop. The analysis predicts migration towards the flow centerline.
\end{abstract}

\maketitle

\section {Introduction}

In his pioneering studies of blood flow, Poiseuille observed the motion of cells
away from the walls of small vessels \cite{Sutera--Skalak:1993}. This phenomenon
proved to be a universal feature of particulate flows in tubes, and the
mechanisms leading to particle motion transverse to the flow direction have
since attracted considerable research effort \cite{Leal:1980}.

A classic result in microhydrodynamics is that lateral migration of a
neutrally-buoyant, non-deformable spherical particle is prohibited in the
creeping-flow limit, where viscosity dampens fluid acceleration and inertial
effects are negligible.  The result arises from the linearity of the Stokes
equations and boundary conditions, and the symmetry of the problem under
flow-reversal \cite{Bretherton:1962, Lealbook}.  However, such cross-stream
drift may occur if the symmetry is lost, \emph{e.g.}\ by particle deformation in
a shear gradient or in the presence of a wall \cite{Leal:1980}.  A
small-deformation solution for a drop with a clean (surfactant-free) interface
predicts lateral motion in unbounded Poiseuille flow, with the direction of
motion depending on the ratio of drop and suspending fluid viscosities
\cite{Chan-Leal:1979}.  In this note, we investigate the possibility of
cross-stream migration of a non-deforming \emph{spherical} drop, induced by
asymmetric interfacial tension resulting from redistribution of a surfactant.

\section{Problem formulation}

Let us consider a drop with radius $a$ and viscosity $\visrat\eta$, embedded in
an ambient fluid with viscosity $\eta$.  A surfactant, insoluble in the bulk
phases, is adsorbed on the drop interface; diffusion of surfactant is neglected.
The average surfactant concentration is $\Gamma_\eq$, the corresponding
interfacial tension is $\sigma_\eq$, and $\sigma_0$ is the interfacial tension
in the absence of surfactant.  The drop is placed in an unbounded plane Poiseuille flow
$\bv^\infty =\left(U'-\alpha y'^2\right) \uvc{x},$ where $\alpha$ is a measure
of the curvature of the flow profile, $\uvc{x}$ is the unit-vector in the flow
direction, and $U'\uvc{x}$ is the velocity at the centerline. If the initial drop
position is at a distance $y_0$ off-center, the drop ``sees'' a combination of
quadratic, linear shear, and uniform flows.  In a
coordinate system centered on, and translating with, the drop, the flow becomes
\begin{equation}
\label{inf_flow}
\bv^\infty=\left(-\dot\gamma y-\alpha y^2\right)\uvc{x} -\bU_\mig\,,
\end{equation}
where $\dot\gamma=2\alpha y_0$ is the local shear rate,
 and the migration velocity
$\bU_{\mig}$ is the difference between the
velocities of the drop and the undisturbed flow at the drop center.

\begin{figure}
\includegraphics*[width=2in]{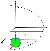}
\caption{An illustration of the problem.
A drop is placed  in unbounded Poiseuille flow at a distance $y_0$ from the flow centerline.}
\label{fig1}
\end{figure}

In creeping flows, a drop remains spherical provided that the capillary number
is small
\begin{equation}
\label{capillary number}
\Ca=\frac{\eta \alpha a^2}{\sigma_\eq}\ll 1\,,
\end{equation}
where the velocity scale is $U_c=\alpha a^2$.
Convection by the surrounding flow creates non-uniformities in the surfactant
distribution.  The ratio of surfactant relaxation and convection time scales defines the inverse
Marangoni number
\begin{equation}
\label{inverse Marangoni number}
\Ma^{-1}=\frac{\eta \alpha a^2}{\Delta\sigma} \; , \;
\Delta\sigma\!=\!\sigma_0\!-\!\sigma_\eq\,,
\end{equation}
which can also be viewed as the ratio of viscous stresses to the characteristic Marangoni stresses
(gradients in surface tension). 
For a dilute surfactant monolayer,
a linear ``perfect gas'' interfacial
equation of state relates   surface tension and local surfactant concentration:
$\sigma(\Gamma)-\sigma_\eq=\Ma \left(1-\Gamma/\Gamma_\eq\right)$.
Henceforward, all quantities are normalized using $a$, $\eta$, $U_c$, and
$\Gamma_\eq$.

The flow 
is described by the Stokes equations.
The velocity and pressure fields $\bv $ and $p$ satisfy
\begin{equation}
\begin{split}
  \nabla p &= \nabla^2 \bv \quad ,\quad     \nabla\cdot\bv = 0 \quad \mathrm{outside} \, , \\
  \nabla p &= \lambda\nabla^2 \bv \quad , \quad     \nabla\cdot\bv = 0  \quad \mathrm{inside} \, ,
  \label{eq:floweqs}
\end{split}
\end{equation}
and $\bv$ is continuous everywhere.
The viscous shearing stresses at the interface are balanced by Marangoni
stresses $\nabla_s \sigma=-\Ma\nabla_s\Gamma$. To close the problem, we need an
evolution equation for the surfactant. This is supplied by a
conservation equation for bulk-insoluble, non-diffusing surfactant on a moving interface:
\begin{equation}
\label{surfeq}
\frac{\partial \Gamma}{\partial t}+\nabla_s\cdot(\bv_s\Gamma)+\Gamma
(\bv\cdot\uvc{n})\nabla\cdot\uvc{n}=0 \, ,
\end{equation}
where $\bv_s$ is the velocity component tangential to the surface.  For a
sphere, the mean curvature $\nabla\cdot\uvc{n}=2$\,.

\section{Solution}

By the linearity of the Stokes equations, the perturbation in the imposed
Poiseuille flow around the surfactant-covered drop may be given as a
superposition of two components. The first is a flow about a clean, surfactant--free drop placed
in the Poiseuille flow. The second is a flow driven by Marangoni stresses
arising from non-uniform surfactant coverage of a drop in a quiescent fluid
\cite{Blawzdziewicz-Vlahovska-Loewenberg:2000}.  In similar fashion, the drop
migration may be decomposed into
\begin{equation}
\label{migration vel2}
\bU_{\mig} =\bU^\cd_{\mig}+ \bU^\surf_{\mig}(\Gamma)\,.
\end{equation}
The
velocity perturbation due to a clean spherical drop gives rise to slip, but no
lateral migration
\cite{Nadim-Stone:1991}:
\begin{equation}
\label{migration vel:cd}
\bU^\cd_{\mig}= -\frac{\lambda}{3\lambda+2}\uvc{x}\,.
\end{equation}
Note that $\bU^\cd_{\mig}$ is normalized by $\alpha a^2$, so the dimensional
slip velocity depends on the flow curvature.
To determine the surfactant contribution $\bU^\surf_{\mig}$,
we solve for the
velocity field about the surfactant-covered drop using the formalism
developed by B{\l}awzdziewicz \emph{et al.}\
\cite{Blawzdziewicz-Vlahovska-Loewenberg:2000}.
In the original work, the method was applied to the dynamics of a stationary
surfactant-covered drop in a linear flow.  We have generalized the approach to
treat a translating drop in higher-order flows.
Owing to the spherical symmetry of the problem, all quantities are represented
in terms of spherical harmonics (see Appendix~\ref{Ap:vel fields} for definitions). Accordingly, the
surfactant concentration $\Gamma$ is expanded in scalar harmonics \refeq{spher harmonics}
\begin{equation}
\label{Gamma expanded in harmonics}
\Gamma=1+\sum_{j=1}^\infty \sum_{m=-j}^{j} g_{jm}Y_{jm}\left(\theta,
\varphi\right)\,,
\end{equation}
 where   $\theta,\, \varphi$ denote the spherical coordinate angles.
The velocity field is expanded in   a set of fundamental solutions of the Stokes
equations \refeq{vel basis -} and \refeq{vel basis +} \cite{Cichocki-Felderhof-Schmitz:1988, Blawzdziewicz-Vlahovska-Loewenberg:2000}
\begin{equation}
\label{vel}
\begin{split}
\bv_{\out}=&\;
c^\infty_{jmq}\left[\bu^+_{jmq}(\br)-\bu^-_{jmq}(\br)\right]+c_{jmq}\bu^-_{jmq}(
\br)\,,\\
\bv_{\ins}=&\; c_{jmq}\bu^+_{jmq}(\br)\,.\\
\end{split}
\end{equation}
Summation over repeated indices is implied.   The functions
$\bu^{\pm}_{jmq}$ are vector solid harmonics
 related to the harmonics in the  Lamb solution.  With respect to a sphere,
$\bu^{\pm}_{jm2}$
is radial, while $\bu^{\pm}_{jm0}$  and $\bu^{\pm}_{jm1}$ are tangential;
$\bu^{\pm}_{jm1}$ is surface-solenoidal ($\nabla_s \cdot
\bu^{\pm}_{jm1}=0$).
The far-field (imposed) flow is specified by $\bv^\infty \!=\!
c^\infty_{jmq}\bu^+_{jmq}$ ; coefficients for the Poiseuille flow
\refeq{inf_flow} are listed in Appendix \ref{Ap:flows}.  The velocity fields
defined in \refeq{vel} are naturally continuous across the interface because the
basis fields $\bu^{\pm}_{jmq}$ reduce to the corresponding vector spherical
harmonics $\by_{jmq}$ at $r\!=\!1$.  The velocity coefficients $c_{jmq}$ are
determined from the stress balance equations listed in Appendix
\ref{Ap:tractions}.  A fixed spherical shape limits the normal surface velocity
to
rigid body translation.  Thus, for $j \!>\! 1$, $c_{jm2}\!=\!0$, and the other
two coefficients
$c_{jmq}$ are determined from the tangential stress balances alone, as use of
the normal stress balance in conjunction with shape specification
over-constrains the problem.  However, both the tangential and normal stress
balances are
used to determine the $c_{1mq}$, which correspond to translational and
rotational motions of the drop.

The velocity coefficients are
decomposed into clean-drop and surfactant contributions:
$c_{jmq}\!=\!c^\cd_{jmq} + c^\surf_{jmq}(\Gamma)$.
We obtain for the surfactant-driven flow
\begin{equation}
\label{surfactant vel}
	\begin{split}
		c^\surf_{jm0}=& -\frac{\delta_{1j}}{(3
\visrat+2)}\,\frac{\sqrt{2}}{\;3}\Ma\,g_{jm} \\
&-
\frac{(1-\delta_{1j})}{\visrat+1}\frac{\sqrt{j(j+1)}}{(2j+1)}\Ma\,g_{jm} \,,\\
		c^\surf_{jm1}=&\;\, 0 \,,\\
		c^\surf_{jm2}=&\quad\, \frac{\delta_{1j}}{(3
\visrat+2)}\,\frac{2}{3}\Ma\,g_{jm} \,,
	\end{split}
\end{equation}
where $\delta_{kl}$ is the Kronecker delta. The solution for $j \!>\! 1$ is
identical to that of B{\l}awzdziewicz \emph{et al.}\
\cite{Blawzdziewicz-Vlahovska-Loewenberg:2000}.  The clean-drop flow is
\begin{equation}
\label{clean drop vel}
	\begin{split}
		c^\cd_{jm0}=&\, \frac{\delta_{1j}}{3 \visrat+2}\left[(2\visrat+3)c^\infty_{jm0} + \sqrt{2}(\visrat-1)c^\infty_{jm2}\right] \\
+&\frac{1-\delta_{1j}}{\visrat+1}\left[2c^\infty_{jm0}-\frac{3}{\sqrt{j(j+1)}}c^\infty_{jm2}\right] \,,\\
		c^\cd_{jm1}=&\, \frac{2j+1}{2+j+\visrat(j-1)} c^\infty_{jm1} \,,\\
		c^\cd_{jm2}=&\, \frac{\delta_{1j}}{3 \visrat+2}\left[\sqrt{2}(
\visrat-1)c^\infty_{jm0} + ( \visrat+4)c^\infty_{jm2}\right] \,.
	\end{split}
\end{equation}

The drop migration velocity is the difference between the volume-averaged
velocity of the drop and the undisturbed velocity at the drop center
\begin{equation}
\label{migration vel3}
	\bU_{\mig}= -\bv^\infty(0) + \frac{3}{4\pi} \int_{drop} \bv(\br) \dif \br\,.
\end{equation}
For a spherical drop, only the $\bu^{+}_{1m2}$ contribute to the integral. 
Hence we obtain
\begin{equation}
\begin{split}
\bU_{\mig}=\textstyle{\sqrt{\frac{3}{8\pi}}}&\left[-(c_{112}-c_{1-12})\uvc{x} \right.\\
&\left.-\im (c_{112}+c_{1-12})\uvc{y}+\textstyle{\sqrt{2}}\,c_{102}\uvc{z}
\right]\,.
\end{split}
\end{equation}
The above expression, in conjunction with \refeq{clean drop vel} and \refeq{cinf
plane poise}, yields the slip term \refeq{migration vel:cd}.  Using
\refeq{surfactant vel} instead of \refeq{clean drop vel} yields the
surfactant-induced migration
\begin{equation}
\label{surfmig}
\begin{split}
\bU_{\mig}^\surf =
\textstyle{\frac{\Ma}{\sqrt{6\pi}(3\lambda+2)}}&\left[-(g_{11}-g_{1-1}) \uvc{x}
\right.\\
&\left.-\im (g_{11}+g_{1-1})\uvc{y}+\textstyle{\sqrt{2}}\,g_{10}\uvc{z}
\right]\,.
\end{split}
\end{equation}
The surfactant distribution coefficients $g_{1m}$ are determined from the
evolution equation \cite{Blawzdziewicz-Vlahovska-Loewenberg:2000,
Vlahovska-Blawzdziewicz-Loewenberg:2002}
\begin{equation}
\label{evolution for gamma}
\begin{split}
\frac{\partial g_{jm}}{\partial t}&=
  C_{jm}+\left[\Omega_{jmj_2m_2}+\Lambda_{jmj_2m_2}\right]g_{j_2m_2}\\
  &
  +\Ma \left[W(j)g_{jm}+ \Theta_{jmj_1m_1j_2m_2}g_{j_2m_2}g_{j_1m_1}\right]\,.
       \end{split}
\end{equation}
The terms in this equation are defined in Appendix \ref{Ap:evolution}.  The
first three terms describe convection of surfactant by the imposed flow: $\Omega$
describes rotation by the linear shear component of the flow, while  $C_{jm}$ and $\Lambda$ describe  redistribution by other components of the flow. The terms proportional to $\Ma$
pertain to  flows driven by Marangoni stresses. The linear term
describes relaxation towards the equilibrium uniform surfactant distribution,
while the quadratic term describes convection of surfactant by the
surfactant-induced flow.

The migration velocity of a spherical
drop is related to the $j\!=\!1$ modes of the surfactant
distribution \refeq{surfmig}.  Equation \refeq{evolution for gamma} shows that these modes can be created by several
interactions between velocity and surfactant field components, \emph{e.g.}\ quadratic flow ($j\!=\!3$) coupling to sheared surfactant ($j\!=\!2$).

For arbitrary distortions of the surfactant concentration, equation
\refeq{evolution for gamma} must be integrated numerically to determine $g_{1m}$ and $\bU_\mig$. Analytical solutions are possible if we consider quasi-steady,
slightly-perturbed surfactant distributions. Quasi-steady implies that the
surfactant evolution occurs on a faster
time scale than the drop migration, \emph{i.e.} $\eta a/\Delta \sigma\ll {a}/{U_\mig}$.
Small disturbances in the surfactant
concentration admit a solution in
the form of a regular perturbation expansion
\begin{equation}
\bU_\mig=\bU_\mig^{(0)}+\bU^{(1)}_\mig+\ldots
\end{equation}
The choice of small parameter depends on the flow regime of interest.  In weak
flows, surfactant relaxation towards the equilibrium distribution is fast, and
$\Ma^{-1}$ is the relevant small parameter, being the ratio of the time scales associated
with Marangoni relaxation and distortion by convection. If the drop is far away
from the centerline ($y_0\!\gg\!1\,$), the shear flow $ 2\alpha
y_0 y\,$ is dominant over the quadratic flow $\alpha y^2$. The extensional part of the shear convects surfactant towards two poles on the drop that correspond to the straining axis of the flow.  Concurrently, the rotational component of the shear rotates the drop, limiting these distortions.  The ratio of
rotational and extensional time scales \cite{Vlahovska-Blawzdziewicz-Loewenberg:2002} is $\sim \!\visrat^{-1}$, so this is an appropriate small parameter for very viscous drops. Next, we examine these two regimes in more detail.

\section{Results and Discussion}

{\subsection{Weak flow / Nearly-incompressible surfactant: $Ma^{-1}\ll1$ and
$\visrat= O(1)$}}

In this regime, only Marangoni stresses oppose the convection of surfactant.
If $\Ma^{-1}\ll 1$, following the discussion in B{\l}awzdziewicz \emph{et al.}\
\cite{Blawzdziewicz-Vlahovska-Loewenberg:2000}, we introduce a
regular expansion for the surfactant concentration:
\mbox{$g_{jm}=\sum_{k=0}^\infty
\Ma^{-k-1} g_{jm}^{(k)}\,$}.
At leading order, the evolution equation \refeq{evolution for gamma} becomes
\mbox{$0=C_{jm}+{W(j)} {g}^{(0)}_{jm}$}. Solving for ${g}^{(0)}_{jm}$ yields
\begin{eqnarray}
\label{largeMa gamma0}
	{g}^{(0)}_{1m} &= &\frac{5}{\sqrt{2}}\left(c^{\infty}_{1m0} - \sqrt{2}
\,c^{\infty}_{1m2}\right)\quad \mbox{and}\\
 g^{(0)}_{jm}&=&\frac{2j+1}{j(j+1)}\left[2\sqrt{j(j+1)}c^{\infty}_{jm0}-
3c^{\infty}_{jm2}\right]\,\mbox{for}\,\, j>1 \,. \nonumber
\end{eqnarray}
Inserting the ${g}^{(0)}_{1m}$ expression in \refeq{surfmig} and
\refeq{migration vel2} gives
\begin{equation}
\label{Faxen result}
\bU^{(0)}_{\mig} =-\frac{1}{3}\uvc{x} \,.
\end{equation}
Thus, at leading order, the Marangoni stresses immobilize the interface. The
surface flow is
incompressible, and the drop behaves like a rigid sphere.

At next order, the evolution equation is
\begin{equation}
\label{LargeMa evol1}
	\begin{split}
0=&\left[\Omega_{jmj_2m_2}+\Lambda_{jmj_2m_2}\right]{g}^{(0)}_{j
_2m_2} + {W(j)}{g}^{(1)}_{jm} \, \\
&+
\Theta_{jmj_1m_1j_2m_2}{g}^{(0)}_{j_1m_1}{g}^{(0)}_{j_2m_2} \, ,
	\end{split}
\end{equation}
and the ${g}^{(1)}_{1m}$ give rise to a cross-stream migration velocity
\begin{equation}
\label{migVMa}
	\begin{split}
		\bU^{(1)}_{\mig} &= - \Ma^{-1}{ y_0}
\frac{3\lambda+17}{9(\lambda+4)}\,\uvc{y} \, .\\
	\end{split}
\end{equation}
Note that the dimensional cross-stream velocity is quadratic in $\alpha$.
Hence, its direction does not change upon reversal of the flow direction.  Its
magnitude depends linearly on $y_0$, the distance from the centerline, and thus
monotonically on the local shear rate, $\sim \alpha y_0$.
It is directed towards the centerline or, equivalently, towards lower shear
rates, for all values of the
viscosity ratio $\visrat$.

{\subsection{Far from the centerline / High viscosity drops: $\visrat^{-1}\ll1$
and $\Ma=O(\lambda)$}}

In this regime, both rotation and Marangoni relaxation limit convective
distortions of the surfactant distribution.  If $\visrat^{-1}\ll 1$, we expand
\mbox{$g_{jm}= \sum_{k=0}^\infty \visrat^{-k-1}g_{jm}^{(k)}$} and introduce \mbox{$\tilde\Ma \equiv\visrat^{-1}{\Ma}$}.  The leading order equation for the surfactant distribution
is the $\lambda \!\rightarrow\! \infty$ limit of \refeq{evolution for gamma}.
Only the first convection term, the linear Marangoni term, and the rigid
rotation within the $\Omega$ term survive.  Thus,
\begin{equation}
\label{Highvisc evol1}
	\begin{split}
0=\tilde C_{jm} +
\Omega_{jmjm}{g}^{(0)}_{jm}+\tilde{\Ma}\,\tilde{W}(j){g}^{(0)}_{jm}\,,
	\end{split}
\end{equation}
where ${C}(jm)\equiv \lambda^{-1}\tilde{C}(jm)$  and
${W}(j)\equiv\lambda^{-1}\tilde{W}(j)$ as $\lambda \!\rightarrow\! \infty$.
 Solving for $g_{1m}^{(0)}$ yields
\begin{equation}
g_{10}^{(0)} = 0 \quad , \quad g_{1\pm1}^{(0)}=\frac{2}{\mp 2 \tilde{\Ma}+3\im
y_0}\sqrt{\frac{2 \pi}{3}} \, .
\end{equation}
Inserting in \refeq{surfmig} and
\refeq{migration vel2} and keeping terms up to order $\lambda^{-1}$ gives a slip
velocity
\begin{equation}
		 U^x_{\mig}
=-\frac{1}{3}+\frac{2}{9\visrat}-\frac{8}{9\visrat }\frac{
\tilde{\Ma}^2}{9
y_0^2 + 4\tilde{\Ma}^2} \quad ,
\end{equation}
and a cross-stream drift
\begin{equation}
\label{migVLam}
		 U^y_{\mig}=
-\frac{4}{3\visrat}\frac{y_0 \tilde{\Ma}}{9
y_0^2 + 4\tilde{\Ma}^2} \quad .
\end{equation}
As before, reversing the flow direction changes the sign of the slip velocity
but leaves the cross-stream velocity unaffected.
As $\tilde{\Ma}\rightarrow \infty$, $\lambda \rightarrow \infty$, or $y_0 \rightarrow 0$, rigid sphere behavior is recovered; cross-stream motion
vanishes, and the slip is given by the Fax{\'{e}}n result \refeq{Faxen result}.
While the cross-stream velocity is always directed towards the
centerline, its magnitude is a non-monotonic function of position, with
maxima at $y_0 = \pm \frac{2}{3}{\tilde{\Ma}}$.
\\
{\subsection{Arbitrary perturbation in the surfactant distribution}}

The present analysis is easily extended to larger redistributions and the
general surfactant dynamics of the full equation \refeq{evolution for gamma}.  A
drop trajectory $y_0(t)$ is determined by numerical solution of this
equation to obtain  \mbox{$dy_0/dt=U^y_{\mig}\left[y_0, \Gamma(y_0)\right]$}.
The evolution equation for the surfactant distribution was truncated at  $j\!=\!6$\, ; this was enough to produce convergent results.

\begin{figure}[h]
\hspace{-3ex}
\includegraphics*[width=3.5in]{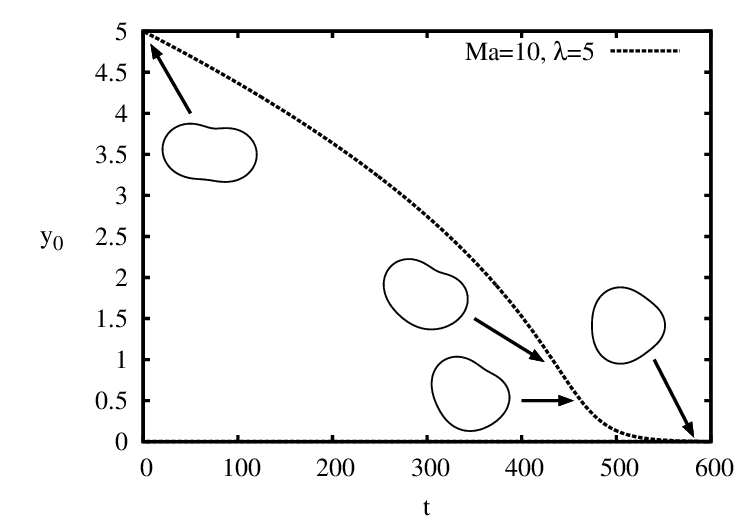}
\vspace{-5ex}
\caption{Drop trajectory (numerical solution for $\Ma=10$, $\visrat=5$) and snapshots of the surfactant distribution in the $xy$ plane (obtained from the large $\lambda$ expansion) at several distances from the centerline.}
\label{typicaltraj}
\end{figure}

\begin{figure}[h]
\hspace{-3ex}
\includegraphics*[width=3.5in]{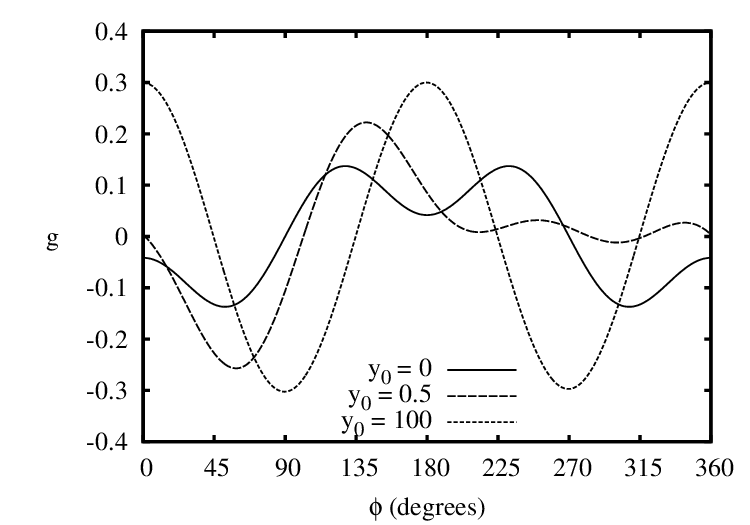}
\vspace{-5ex}
\caption{Distribution in the $xy$ plane of $g= \Gamma - 1$, the deviation from uniform surfactant concentration, for the large $\lambda$ expansion at three distances $y_0$ from the centerline.}
\label{surfdist}
\end{figure}

A typical drop trajectory is shown in Figure \ref{typicaltraj}, along with shapes representing the distribution of surfactant in the $xy$ plane for the high viscosity expansion at several distances from the centerline.  Figure \ref{surfdist} shows three such distributions as a function of the azimuthal angle $\varphi$.  The distribution is three-lobed near the centerline and two-lobed far from the centerline.  These shapes reflect the dominance of the quadratic \mbox{($j\!=\!3$)} and linear shear \mbox{($j\!=\!2$)} components of the flow at these respective locations.  The behavior of the particle is governed by
rotation in the high-shear-rate region far from the centerline, and by Marangoni
effects near the centerline. 

\begin{figure}[h]
\hspace{-3ex}
\includegraphics*[width=3.5in]{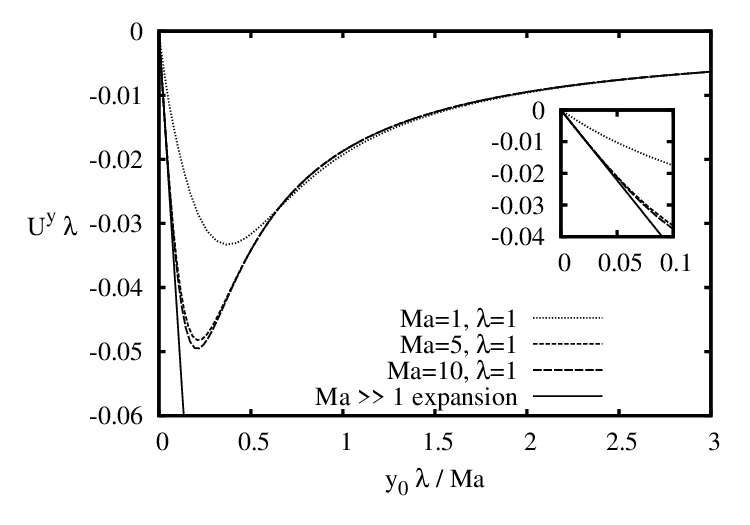}
\vspace{-5ex}
\caption{Rescaled cross-stream migration velocity $U^y_{\mig}$ as a function of rescaled distance from the centerline $y_0$:  Large $\Ma$ expansion and numerical solutions for several values of $\Ma$.}
\label{velvaryma}
\end{figure}

\begin{figure}[h]
\hspace{-3ex}
\includegraphics*[width=3.5in]{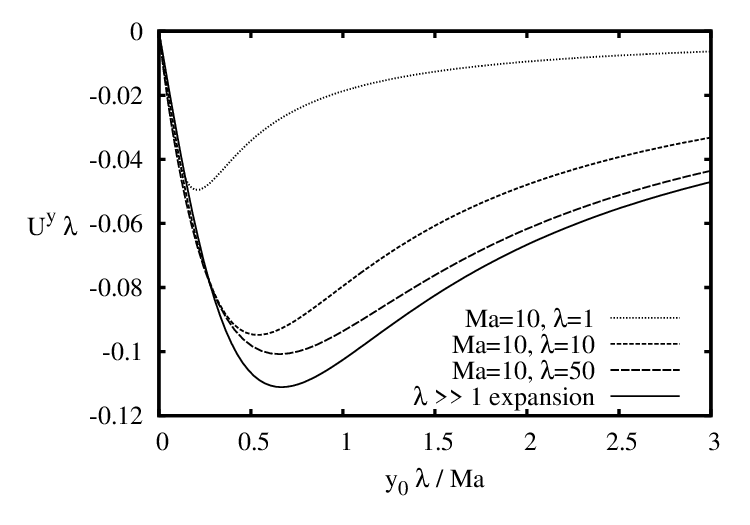}
\vspace{-5ex}
\caption{Rescaled cross-stream migration velocity $U^y_{\mig}$ as a function of rescaled distance from the centerline $y_0$:  Large $\lambda$ expansion and numerical solutions for several values of $\lambda$.}
\label{velvaryvisc}
\end{figure}

The $g_{1m}$ are linear gradients in surfactant concentration; migration towards the centerline results from an excess of surfactant on the drop hemisphere facing the centerline.  At the centerline, $g(\varphi) \!=\! g(-\varphi)$ and $U^y_{\mig}$ vanishes.  This symmetry is also approached as $y_0 \rightarrow \infty$ in the unbounded flow.  Thus, $U^y_{\mig}$ has an extremum, and the trajectory an inflection point.  This point is a rough indicator of the distance from the centerline, of order $\frac{\Ma}{\lambda}$, beyond which the large Marangoni expansion is inadequate even as a qualitative descriptor of the drop's behavior.  The high viscosity expansion captures the qualitative behavior of quasi-steady trajectories at all distances.  Figures \ref{velvaryma} and \ref{velvaryvisc} show the cross-stream migration velocity as a function of distance from the centerline for several values of the parameters, along with the two expansions.  

The numerical results suggest that cross-stream migration of a
surfactant-covered spherical drop is directed towards an equilibrium position at the flow centerline.  
This result agrees with
those found for other particles, such as capsules \cite{Helmy} and vesicles
\cite{Danker-Vlahovska-Misbah}, whose interfaces are governed by Marangoni-like
stresses, and with simulations of surfactant-covered, deformable, equi-viscous
drops \cite{Janssen-Anderson:2008}.

\begin{figure}[h]
\hspace{-3ex}
\includegraphics*[width=3.5in]{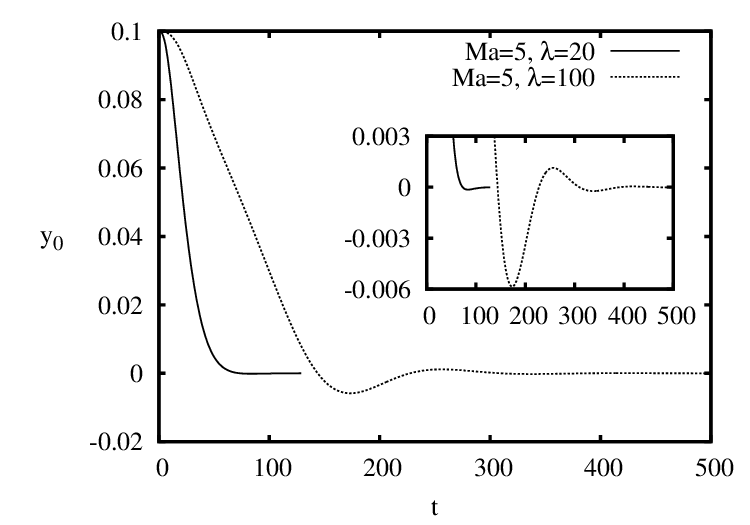}
\vspace{-5ex}
\caption{Two drop trajectories that overshoot and return to the centerline.}
\label{oscillations}
\end{figure}

However, a quasi-steady treatment of the drop dynamics is not always appropriate, and more complicated non-monotonic trajectories are possible.  We note one interesting feature that occurs
when $\frac{\Ma}{\lambda}$ is small and the drop experiences its greatest cross-stream velocity very close to the centerline.  In such cases, the drop may overshoot and execute small-amplitude damped oscillations around the centerline, as shown in Figure \ref{oscillations}.  Similar behavior has been predicted for vesicles \cite{Danker-Vlahovska-Misbah}.  

\begin{figure}[h]
\hspace{-3ex}
\includegraphics*[width=3.5in]{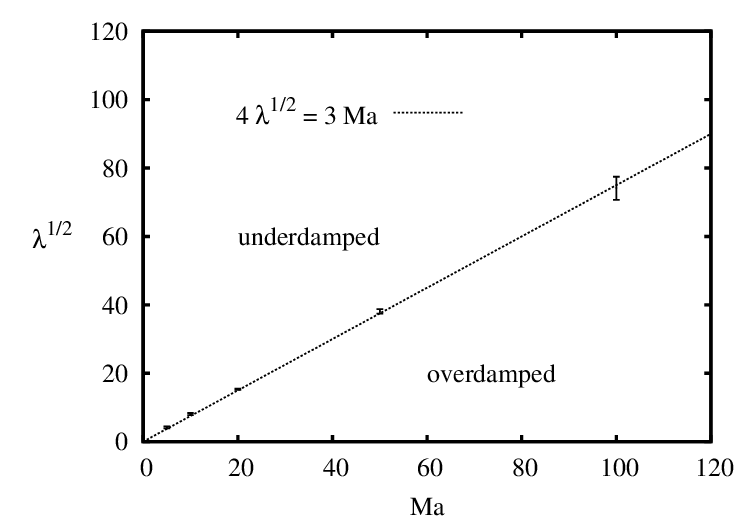}
\vspace{-5ex}
\caption{Empirically determined critical damping values of $\sqrt{\lambda}$ for $5 \le \Ma \le 100$ ; the line $\sqrt{\lambda}=\frac{3}{4}\,\Ma$ is an approximate fit.}
\label{critical}
\end{figure}

We may consider a linearized dynamics for a viscous drop near the centerline by noting that \mbox{$\dot{y}_0 \sim -\frac{\Ma}{\lambda}(g_{11}+g_{1-1})$} according to \refeq{surfmig}.  The damping term is provided by the linear relaxation term of \refeq{evolution for gamma}, which \mbox{$\sim -\frac{\Ma}{\lambda}$}.  Finally, either \refeq{migVMa} or \refeq{migVLam} indicates that when \mbox{$\ddot{y}_0 \rightarrow 0$} and \mbox{$y_0 \ll 1$}, \mbox{$\dot{y}_0 \sim -\frac{1}{\Ma}y_0$}.  Hence, denoting unknown constants as $k_i$,
\begin{equation}
	\ddot{y}_0 + k_1 \frac{\Ma}{\lambda} \dot{y}_0 + k_2 \frac{1}{\lambda} y_0 = 0 \, .
\end{equation}
Such an oscillator will be critically damped when \mbox{$\Ma= k_3\sqrt{\lambda}$} ; this prediction is supported by the numerical results shown in Figure \ref{critical}.  

\section{Conclusions}
We have shown that the presence of small amounts of surfactant
can significantly affect drop motions in quadratic flows.  In contrast with a
clean, deformable drop \cite{Chan-Leal:1979}, for which the direction of
migration depends on viscosity ratio, a surfactant-covered spherical drop always
migrates towards the flow centerline.

\section{Acknowledgments}
This work was supported in part by NSF CAREER award CBET--0846247. Acknowledgment is made to the Donors of the American Chemical Society Petroleum Research Fund for  partial support of this research.

\appendix

\section{Harmonics and velocity fields}
\label{Ap:vel fields}

Scalar and vector spherical harmonics are defined as
\begin{equation}
\label{spher harmonics}
	\begin{split}
		Y_{jm}(\theta,\varphi) &= \textstyle \left[\frac{2j+1}{4\pi}
\frac{(j-m)!}{(j+m)!}\right]^\frac{1}{2} (-1)^m P_j^m(\cos\theta)e^{{\rm
i}m\varphi} \, , \\
		&\by_{jm0}=\left[j\left(j+1\right)\right]^{-\frac{1}{2}}r \, \nabla_{\Omega}
Y_{jm} \, , \\
		&\by_{jm1}=-\im \uvc{r} \times \by_{jm0} \, , \\	
		&\by_{jm2}=\uvc{r} \, Y_{jm} \, , \\
	\end{split}
\end{equation}
where the $P_j^m$ are the Legendre polynomials, and $\nabla_{\Omega}$ is the
angular part of the gradient operator.  The velocity basis functions are
\begin{subequations}
\label{vel basis -}
\begin{align}
\label{-vel 0}
		\bu^-_{jm0}&={\textstyle\frac{1}{2}}r^{-j}\left(2-j+j r^{-2}\right)\by_{jm0}
\nonumber\\ &
+{\textstyle\frac{1}{2}}r^{-j}\left[j\left(j+1\right)\right]^{\frac{1}{2}}\left(
1-r^{-2}\right) \by_{jm2} \, , \\
\label{-vel 1}
		\bu^-_{jm1}&=\textstyle r^{(-j-1)} \by_{jm1} \, , \\
\label{-vel 2}
		\bu^-_{jm2}&={\textstyle\frac{1}{2}}r^{-j}\left(2-j\right)
(\textstyle\frac{j}{j+1})^{\frac{1}{2}}\left(1-r^{-2}\right)\by_{jm0}
\nonumber\\ &
		+{\textstyle\frac{1}{2}}r^{-j}\left(j+(2-j)r^{-2}\right)\by_{jm2} \, ,
\end{align}
\end{subequations}
\vspace{-3ex}
\begin{subequations}
\label{vel basis +}
\begin{align}
\label{+vel 0}
		\bu^+_{jm0}&= {\textstyle\frac{1}{2}}r^{j-1}\left(-(j+1)+(j
+3)r^2\right)\by_{jm0} \nonumber\\ &	
-{\textstyle\frac{1}{2}}r^{j-1}\left[j\left(j+1\right)\right]^{\frac{1}{2}}\left
(1-r^2\right)\by_{jm2} \, , \\
\label{+vel 1}
		\bu^+_{jm1}&=\textstyle r^j \by_{jm1} \, , \\
\label{+vel 2}
\bu^+_{jm2}&={\textstyle\frac{1}{2}}r^{j-1}\left(3+j\right)(\textstyle\frac{j+1}
{j})^{\frac{1}{2}}\left(1-r^2\right)\by_{jm0} \nonumber\\ &
		+{\textstyle\frac{1}{2}}r^{j-1}\left(j +3-(j+1)r^2  \right)\by_{jm2} \, .
\end{align}
\end{subequations}

\section{Imposed flow}
\label{Ap:flows}

The unbounded plane Poiseuille flow \refeq{inf_flow} seen by a migrating
particle is represented as
\begin{equation}
\bv^\infty=c_{jmq}^\infty \bu^+_{jmq} \, ,
\end{equation}
with coefficients
\begin{equation}
\label{cinf plane poise}
	\begin{split}
\textstyle		&c^\infty_{3\pm30}=\textstyle\mp \alpha \sqrt{\frac{4\pi
}{105}}\,,\quad \; c^\infty_{3\pm32}=\mp\alpha\sqrt{\frac{\pi }{35}}\,,\\
\textstyle		&c^\infty_{3\pm10}=\textstyle\mp\alpha \frac{2 }{15}\sqrt{\frac{\pi
}{7}}\,,\quad c^\infty_{3\pm12}=\mp\alpha\frac{1}{5}\sqrt{\frac{\pi }{21}}\,,\\
\textstyle		&c^\infty_{2\pm20}=\textstyle\pm\dot\gamma \im \sqrt{\frac{\pi
}{5}}\, , \quad \;\;\; c^\infty_{2\pm22}=\pm \dot\gamma \im \sqrt{\frac{2 \pi
}{15}}\,, \\
\textstyle		&c^\infty_{2\pm1 1}=\textstyle \alpha\frac{2 }{3}\sqrt{\frac{\pi
}{5}}\, , \quad \;\, c^\infty_{101}=-\im \dot\gamma \sqrt{\frac{2 \pi }{3}}\,,
\\
\textstyle		&\quad c^\infty_{1\pm10}=\textstyle\pm
\left(\alpha\frac{4}{5}+2(U_{\mig}^x \mp\im U_{\mig}^y )\right)\sqrt{\frac{\pi
}{3}}\,, \\
\textstyle		&\quad c^\infty_{1\pm12}=\textstyle\pm
\left(\alpha\frac{1}{5}+(U_{\mig}^x \mp\im U_{\mig}^y )\right) \sqrt{\frac{2 \pi
}{3}}\,,\\
\textstyle		&c^\infty_{100}=\textstyle -2U_{\mig}^z \sqrt{\frac{2\pi
}{3}}\,,\quad \;\;
c^\infty_{102}=-2U_{\mig}^z\sqrt{\frac{\pi }{3}}\,.\\
	\end{split}
\end{equation}

\section{Hydrodynamic tractions and stress balances}
\label{Ap:tractions}

Neglecting isotropic contributions, the hydrodynamic tractions on a sphere  may
be expanded in
harmonics $\bt_{\mathrm{hyd}} = \tau_{jmq}\by_{jmq}$, with
\begin{equation}
\label{traction coeff}
	\begin{split}
\tau^{\mathrm{out}}_{jmq}&=\textstyle\sum_{q'}c^\infty_{jmq'}\left(T_{q'q}^{+}-T
_{q'q}^{-}\right)+c_{jmq'}T_{q'q}^{-} \, , \\
		\tau^{\mathrm{in}}_{jmq}&=\visrat \textstyle\sum_{q'}c_{jmq'}T_{q'q}^{+}\, ,
	\end{split}
\end{equation}
where $T^\pm_{q'q}=T^\pm_{qq'}$, $T^\pm_{00} = \pm(2j+1)$, $T^\pm_{02}=
\mp3\left(\frac{j+1}{j}\right)^{\pm\frac{1}{2}}$, $T^\pm_{11} =
\pm\left(j+\frac{1}{2}\right)-\frac{3}{2}$, $T^\pm_{22}=
\pm(2j+1)+\frac{3}{\pm\left(j+\frac{1}{2}\right)-\frac{1}{2}}$, and
$T^\pm_{01}=T^\pm_{12}=0$.

The tangential stress boundary conditions are
\begin{equation}
\begin{split}
\tau_{jm0}^\out-\tau_{jm0}^\ins=&\; \Ma \sqrt{j(j+1)}\,g_{jm}\,,\\
\tau_{jm1}^\out-\tau_{jm1}^\ins=&\; 0 \, ,
\end{split}
\end{equation}
where a linear equation of state relating surface tension and local surfactant
concentration is assumed.
The normal stress balance for a spherical drop requires
\begin{equation}
\tau_{jm2}^\out-\tau_{jm2}^\ins=-2 \Ma \, g_{jm}
\end{equation}
For $j\!>\!1$, we set $c_{jm2}\!=\!0$ and do not use the
normal stress balance condition.

\section{Surfactant evolution equation}
\label{Ap:evolution}
The surfactant conservation equation on a moving interface is given by
\refeq{surfeq}.  Representing all
quantities as harmonics and decomposing the velocity field into clean-drop and
surfactant contributions leads to \refeq{evolution for gamma}.  The clean-drop
terms are
\begin{equation}
\begin{split}
C_{jm}=&\, \left[j(j+1)\right]^{\frac{1}{2}}c^\cd_{jm0}-2c^\cd_{jm2} \, , \\
\Omega_{jmj_2m_2}=&\,
\left[j(j+1)\right]^{\frac{1}{2}}c^\cd_{j_1m_11}\,{\cal{C}}^{01} \, , \\
\Lambda_{jmj_2m_2}=&\,
\left[j(j+1)\right]^{\frac{1}{2}}c^\cd_{j_1m_10}\,{\cal{C}}^{00}
-2\,c^\cd_{j_1m_12}\,{\cal{C}}^{22} \, .
\end{split}
\end{equation}\\
Summation over $j_1$ and $m_1$ is implied, and the $c^0_{jmq}$ coefficients
refer to equations \refeq{clean drop vel}.
Using equations \refeq{surfactant vel} to define the vector $\bar W(j)$ by
\begin{equation}
 c^\surf_{jmq} \equiv \bar W_q(j) \Ma \, g_{jm} \,,
\end{equation}
the surfactant terms are
\begin{eqnarray}
		 {W}(j) \!\!&=&\!\!  \left[j(j+1)\right]^{\frac{1}{2}} \bar W_0(j) -
2\,\bar W_2(j) \,, \\
		 \Theta_{jmj_1m_1j_2m_2} \!\!&=&\!\! \left[j(j+1)\right]^{\frac{1}{2}}
\bar W_0(j_1)\,{\cal{C}}^{00} -2\,\bar W_2(j_1)\,{\cal{C}}^{22} \,. \nonumber
\end{eqnarray}
The Clebsch-Gordan coupling coefficients ${\cal{C}}$ are
\begin{equation}
\begin{split}
		{\cal{C}}^{00} &=\, B\left( \begin{array} {ccc}	j_1& j_2& j\\
												0& 0& 0 \end{array} \right)\\
&\times
\frac{j_1(j_1+1)+j(j+1)-j_2(j_2+1)}{\left [ j(j+1)j_1(j_1+1)
\right]^{\frac{1}{2}}} \,, \\
		{\cal{C}}^{01} &=\, B\left( \begin{array} {ccc}	j_1& j_2& j-1\\
												0& 0& 0 \end{array} \right)\\
&\times\left[\frac{(s+1)(s-2j_2)(s-2j_1)(s-2j+1)}{j(j+1)j_1(j_1+1)}
\right]^{\frac{1}{2}} \,, \\
		{\cal{C}}^{22} &= \, B\left( \begin{array} {ccc}	j_1& j_2& j\\
												0& 0& 0 \end{array} \right) \times 2 \,,\\
\end{split}
\end{equation}
where $s = j+j_1+j_2\,$,
\begin{equation*}
\begin{split}
B = \;&\frac{(-1)^{m}}{2}
\left[\frac{(2j+1)(2j_1+1)(2j_2+1)}{4\pi}\right]^{\frac{1}{2}} \\
&\times\left(
\begin{array} {ccc} j_1& j_2& j\\
													m_1& m_2& -m \end{array} \right )\,,
\end{split}
\end{equation*}
and $ \left( \begin{array} {ccc} j_1& j_2&j\\
						m_1& m_2& m \end{array} \right )$ denotes the Wigner 3$j$-symbol
\cite{Edmonds:1960}.\\

\bibliographystyle{unsrt}

\end{document}